\documentstyle[12pt,epsfig]{article}  
\begin{document} 
\baselineskip=18pt

\def\la{\mathrel{\mathpalette\fun <}}
\def\ga{\mathrel{\mathpalette\fun >}}
\def\fun#1#2{\lower3.6pt\vbox{\baselineskip0pt\lineskip.9pt
\ialign{$\mathsurround=0pt#1\hfil##\hfil$\crcr#2\crcr\sim\crcr}}} 

\begin{titlepage} 
\title{PROBING THE MEDIUM-INDUCED ENERGY LOSS OF BOTTOM QUARKS BY DIMUON
PRODUCTION IN HEAVY ION COLLISIONS AT LHC} 
\vskip 1.5cm 
\author{I.P.Lokhtin and A.M.Snigirev \\ ~ \\ 
\it M.V.Lomonosov Moscow State University, \\  
\it D.V.Skobeltsyn Institute of Nuclear Physics,\\  \it Moscow, Russia } 
\date{}
\maketitle 
\vskip 2 cm   
\begin{abstract} 
The potential of coming experiments on Large Hadron Collider (LHC) to observe the 
rescattering and energy loss of heavy quarks in the dense matter created in heavy ion 
collisions is discussed. We analyze the sensitivity of high-mass $\mu^+\mu^-$ pairs 
from $B \overline{B}$ semileptonic decays and secondary $J/\psi$'s from single $B$ 
decays to the medium-induced bottom quark energy loss.  
\end{abstract}
\end{titlepage}    

\section {Introduction}   

The significant progress in lattice QCD calculations, in particular, 
including dynamical quarks, strongly suggests that the deconfinement of 
hadronic matter and chiral symmetry restoration 
must happen at temperatures above $T_c \sim 200$ MeV~\cite{lattice}. 
Experimental studies of the properties of strongly interacting matter at high
enough energy densities for a relatively long-lived quark-gluon plasma 
(QGP) to be 
formed is one of the goals of modern high energy physics (see, for 
example, reviews in
Ref.~\cite{satz182,muller96,lok_rev,bass_rev}). It is expected that the 
quark-hadron 
phase transition, which likely occurred during the first few microseconds 
of the evolution of the universe, can be attained in heavy-ion accelerators.

In recent years, a great deal of attention has been devoted 
to "hard" probes of the QGP including heavy quarkonia, hard hadrons, jets, 
and high mass dimuons which are not part of the thermalized system, thus
carrying information about the early stages 
of the evolution. In particular, the predicted charmonium suppression by 
screening of bound $c \overline{c}$ pairs ("colour dipole") in 
a plasma~\cite{satz86} or 
dynamical dissociation  by semi-hard deconfined gluons~\cite{khar94} 
is one of the most promising signals of QGP formation. A 
similar phenomenon has been observed in the most 
central Pb+Pb collisions at CERN-SPS~\cite{na50}: the anomalously small 
$J/\psi$ to Drell-Yan ratio, inconsistent with pre-resonance 
absorption in cold 
nuclear matter. Although the interpretation of this phenomenon as a result of 
QGP formation is plausible, alternative explanations such as rescattering 
with comoving 
hadrons can not be fully dismissed~\cite{vogt-psi}. For the heavier 
$b \overline{b}$ bound 
states (the $\Upsilon$ family), a similar suppression effect in QGP 
is expected at the
higher temperatures which can be attained in heavy ion collisions 
at the LHC. 

Along with quarkonium suppression one process of interest 
is the passage of coloured
jets through dense matter. Dijets are created at the very beginning of the 
collision process, $\la 0.01$ fm$/c$, by the 
initial hard parton-parton scatterings. 
These hard partons pass through the dense matter formed by minijet 
production at longer time scales, $\sim 0.1$ fm$/c$, 
and interact strongly with the constituents of 
the medium. The inclusive cross section for hard jet production is 
negligible at the  
SPS but increases rapidly with the collision energy. Thus these jets 
will play an  
important role at RHIC (with Au+Au collisions at $\sqrt{s} = 200 A$ GeV) 
and LHC (with Pb+Pb collisions at $\sqrt{s} = 5.5 A$ TeV). The challenge 
is to determine the behaviour of coloured jets in 
dense matter~\cite{loss_rev} due to coherent medium-induced gluon 
radiation~\cite{ryskin,gyul94,baier,zakharov,urs99,vitev} and collisional 
energy loss due to elastic rescatterings~\cite{bjork82,mrow91,lokhtin1}. 
Since the jet rescattering 
intensity strongly increases with temperature, formation of a "hot" QGP at 
initial temperatures up to $T_0 \sim 1$ GeV at LHC~\cite{eskola94} 
should result in much larger 
parton energy loss compared to "cold" nuclear matter or a hadronic gas. 

The following signals of medium-induced energy loss have been identified 
as being observable in ultrarelativistic heavy ion collisions: \\
1) The suppression of high-$p_T$ jet pairs~\cite{gyul90,lokhtin2} with a 
corresponding enhancement of the monojet-to-dijet 
ratio~\cite{gyulqm95,lokhtin3}. Jets are produced 
in the initial scattering processes such as  
$$gg \rightarrow gg~,~~~~qg \rightarrow qg~,~~~~qq\rightarrow qq~,~~~~~gg
\rightarrow q \overline q~$$  
where the $gg \rightarrow gg$ process is dominant. \\  
2) The $p_T$-imbalance between a produced jet with a gauge boson in 
$\gamma+$jet~\cite{wang96} and $Z+$jet~\cite{kvat95} production, e.g.\ 
$$q g \rightarrow q \gamma~, ~~~~~~q g \rightarrow q Z .$$ 
The $Z$-boson is identified through its decays to muon pairs, 
$Z \rightarrow \mu^+ \mu^-$. \\ 
3) The modification of the high mass dimuon spectra from semileptonic 
$B$ and $D$ meson 
decays due to bottom and charm quark energy loss~\cite{shur97,kamp98,lin98}:  
$$gg \rightarrow b \overline{b}~(c \overline{c}) \rightarrow B \overline{B}~(D 
\overline{D}) \rightarrow \mu ^+ \mu^- X .$$ 

The above phenomena can be experimentally studied in heavy ion 
collisions~\cite{note00-060} with Compact Muon Solenoid (CMS), which is the general 
purpose detector designed to run at the LHC~\cite{cms94}. In our previous 
work~\cite{lokhtin2} we discussed the impact parameter distribution in the hard 
jet production processes ($jet + jet$, $\gamma + jet$ and $Z + jet$ channels) at LHC, 
which can shed a light on the dependence of parton energy losses on distance traversed 
(the coherence pattern of the medium-induced radiation can result in non-trivial
behaviour of such dependence, see~\cite{baier}). The main goal of this paper is to 
analyze the sensitivity of both high-mass $\mu^+\mu^-$ pairs 
from $B \overline{B}$ semileptonic decays and secondary $J/\psi$'s 
from single $B$ decays to $b$-quark energy loss in CMS conditions. Note that the 
ALICE experiment~\cite{alice} will also study dilepton production in 
heavy ion collisions although with a different
rapidity acceptance: 
$|\eta^{\mu}| < 2.4$ for CMS while $|\eta^e| < 0.9$ and 
$2.5 < |\eta^{\mu}| < 4$ for 
ALICE. We believe that dimuon production in combination with 
high-$p_T$ jet production 
by gluon and light quark fragmentation can give important 
information about the 
medium-induced effects for both light and heavy partons 
in heavy ion collisions at the LHC. 

The high-mass dimuon spectra in Pb+Pb collisions at LHC has already 
been estimated in 
the CMS acceptance~\cite{lin98} assuming a constant energy 
loss per unit length of
$dE/dx = 1$ GeV$/$fm. In our paper we consider the dynamical evolution of 
heavy quark energy 
loss and rescattering as a 
function of energy density in an expanding "hot" gluon-dominated plasma.  
We also investigate, for the first time, these effects on
secondary $J/\psi$ production from single $B$-meson decays. 

The outline of the paper is as follows. In Sect.2 we discussed the problem of
medium-induced gluon radiation of massive quark for incoherent and coherent cases. 
A model for rescattering of a heavy quark in dense QCD-medium is described in Sect.3. 
Sect.3 and Sect.4 are dedicated to the analysis of calculation results for high-mass 
dimuon and secondary $J/\Psi$ spectra respectively. The summary can be found in
Sect.5. 

\section {Medium-induced gluon radiation from massive quarks} 

We first discuss the effects of the medium on gluon radiation from 
heavy quarks.
We emphasize that the recent theoretical developments on medium-induced gluon 
radiation~\cite{baier,zakharov} are valid in the coherent domain only in the  
ultrarelativistic limit of the quark momenta. Although some attempts 
have been made to 
calculate medium-induced heavy quark energy loss for quarks of mass 
$M_q$~\cite{thoma98}, a full 
description of the coherent radiation from a massive object still does not 
exist.  There are two extreme limits for energy loss by gluon radiation.
In the low $p_T$ limit, $p_T \la M_q$, medium-induced radiation should be 
suppressed by the mass, while the ultrarelativistic limit, 
$p_T \gg M_q$, corresponds to the radiation spectrum of massless quarks. 

The approach developed in Ref.~\cite{thoma98} 
is based on a factorization of the matrix 
elements into elastic scattering and gluon emission~\cite{gunion} where the
multiplicity distribution of the radiated gluons can be written as  
\begin{equation} 
\frac{dN_g}{d\eta d^2q_T} = \frac{3\alpha_s}{\pi^2} 
\frac{l_T^2}{q_T^2(\vec{q_T}-\vec{l_T})^2} , 
\end{equation}  
where $q=(q_0,\vec{q_T},q_3)$ and $l=(l_0,\vec{l_T},l_3)$ are 
the 4-momenta of the emitted 
and the exchanged gluon respectively and $\eta =(1/2)\ln{[(q_0+q_3)/(q_0-
q_3)]}$ is the 
rapidity of emitted gluon relative to the initial quark  momentum. 
It is valid in a
limited $q_T$-region for small rapidities $\eta \sim 0$ and $l_T(q_0/E) 
\ll q_T$. The
coherent LPM effect is taken into account simply by including a formation time 
restriction via 
a step function~\cite{gyul94}. However, for light quarks, 
it leads to a different dependence of the radiative 
energy loss on the initial quark 
energy $E$ than found in other, more recent works~\cite{baier,zakharov}. 

In our case, the main contribution to high-mass dimuon and secondary $J/\psi$ 
production is due to $b$-quarks with "intermediate" values of 
$p_T \ga 5$ GeV$/c$, 
expected to be rather close to the incoherent regime. In order to estimate the 
sensitivity of the dimuon spectra to medium-induced effects, we consider 
two extreme 
cases: (i) the "minimum" effect with collisional energy loss only and (ii) the
"maximum" effect with collisional and radiative energy loss in 
the incoherent limit 
of independent emissions without taking into account the LPM coherent 
suppression of 
radiation (i.e. $dE/dx \propto E$ and is independent of path length, $L$). 
In the latter scenario we use the Bethe-Heitler cross section obtained in 
relativistic 
kinematics and derive the medium-induced radiative energy loss per unit 
length~\cite{zakharov} as the integral over the gluon radiation spectrum 
\begin{eqnarray}
\label{rad_los}
& &  \frac{dE}{dx} = E \rho \int \limits_0^{1-M_q/E} dy \frac{4\alpha_s 
C_3(y) (4-4y+2y^2)} {9\pi y \left[ M_q^2y^2+m_g^2(1-y)\right] } , \\ 
& & C_3 (y) = \frac{9\pi \alpha_s^2 C_{ab}}{4} \left[ 1+(1-y)^2-y^2\right]  
\ln{\frac{2\left( \alpha_s^2 \rho E y(1-y)\right) ^{1/4}}{\mu_D}} , 
\nonumber     
\end{eqnarray}  
where $m_g \sim 3T$ is the effective mass of the emitted gluon at 
temperature $T$, 
$y = q_0/E$ is the fraction of the initial quark energy 
carried by the emitted gluon, $\rho \propto T^3$ is the density of the 
medium and the 
Debye screening mass squared, $\mu_D^2=4 \pi\alpha_s T^2(1+N_f/6)$, 
regularizes the 
integrated parton rescattering cross section. 

\section {Heavy quark rescattering in a dense QCD-medium} 

We have developed a Monte-Carlo simulation of the mean free 
path of heavy quarks ($M_b=5$ GeV and $M_c=1.5$ GeV) 
in an expanding QGP formed in the 
nuclear overlap zone in Pb+Pb collisions. The details of the 
geometrical model of 
hard quark production and the quark passage through dense matter 
can be found in our 
recent work~\cite{lokhtin2}. In general, the intensity of 
rescattering and energy
loss are sensitive to the initial conditions (energy density $\varepsilon_0$
and formation time $\tau_0$) and space-time evolution of the medium, treated
as a longitudinally expanding quark-gluon fluid. The partons are produced on 
a hyper-surface of equal proper times 
$\tau = \sqrt{t^2 -  z^2}$~\cite{bjorken}. 

If the mean free path of a hard parton is larger than the 
screening radius in the 
QCD-medium, $\lambda \gg \mu_D^{-1}$, 
the successive scatterings can be treated as 
independent~\cite{gyul94}. The transverse distance 
between successive scatterings, 
$\Delta x_i = (\tau_{i+1} - \tau_i) v_T = (\tau_{i+1} - \tau_i) p_T/E$, 
is generated according to the probability density  
\begin{equation} 
\frac{dP}{d(\Delta x_i)} = \frac{1}{\lambda(\tau_{i+1})} 
\exp{(-\int\limits_0^{\Delta x_i} \frac{ds}{\lambda (\tau_i + s)})}
\end{equation}
where the mean free path is 
$\lambda = 1/(\sigma \rho )$. The density of the medium 
$\rho (\tau)$ and the quark rescattering cross section $\sigma (\tau)$ 
are functions 
of proper time. Then the basic kinetic integral equation for the energy loss 
$\Delta E$ as a function of initial energy $E$ and path length $L$ has the 
form 
\begin{eqnarray} 
& & \Delta E (L,E) = \int\limits_0^Ldx\frac{dP(x)}{dx}
\lambda(x)\frac{dE(x,E)}{dx} \, , \\ 
& & \frac{dP(x)}{dx} = \frac{1}{\lambda(x)}\exp{\left( -x/\lambda(x)\right) }
\, , \nonumber 
\end{eqnarray} 
where the current transverse coordinate of a quark, $x(\tau)$, is 
determined from $dx/d\tau = v_T$ with $x = \tau$ at $v_T$ = 1. 

The dominant contribution to the differential cross section $d\sigma / dt$ for 
scattering of a quark with energy $E$ and momentum $p = \sqrt{E^2-M_q^2}$ off 
the ``thermal" partons with energy (or effective mass) $m_0 (\tau) \sim 3T 
(\tau) \ll E$ at 
temperature $T$ can be written in the target frame as~\cite{gyul94,thoma98}  
\begin{equation} 
\frac{d\sigma_{ab}}{dt} \cong C_{ab} \frac{2\pi\alpha_s^2(t)}{t^2} 
\frac{E^2}{p^2}  
\end{equation} 
where $C_{ab} = 9/4$, 1, and 4/9 for $gg$, $gq$ and $qq$ scatterings 
respectively.   
The strong coupling constant is 
\begin{equation} 
\alpha_s (t) = \frac{12\pi}{(33-2N_f)\ln{(t/\Lambda_{\rm QCD}^2)}} \>   
\end{equation} 
for $N_f$ active quark flavours and QCD scale parameter $\Lambda_{\rm QCD}$ 
on the order of 
the critical temperature,  $\Lambda_{\rm QCD}\simeq T_c$. The integrated 
parton scattering 
cross section is regularized by the Debye screening mass squared $\mu_D^2$: 
\begin{equation} 
\sigma_{ab}(\tau) = \int\limits_{\mu^2_D}^{t_{\rm max} }dt
\frac{d\sigma_{ab}}{dt}\>  
\end{equation} 
where $t_{\rm max}=[ s-(M_q+m_0)^2] [ s-(M_q-m_0)^2 ] / s$ and 
$s=2m_0E+m_0^2+M_q^2$. 

In the $i$-th rescattering off a comoving medium constituent (i.e. with the same
longitudinal rapidity $Y$) with squared momentum transfer $t_i$ and effective
mass $m_{0i}$, the quark loses total energy $\Delta e_i$ and transverse energy 
$\Delta e_{T i}$ as well as getting a transverse momentum kick $\Delta 
{\bf k}_{t i}$ relative to the initial momentum $\bf {p_T}$. It is 
straightforward to evaluate $\Delta e_i$, $\Delta e_{T i}$ and 
$\Delta {\bf k}_{t i}$: 
\begin{eqnarray} 
& & \Delta e_{T i} \simeq \frac{t_i}{2m_{0i}} , \\ 
& & \Delta e_i \simeq \Delta e_{T i} \cosh{Y}-\Delta k_{t i}\cos{\phi}\sinh{Y} , \\ 
& & \Delta k_{t i} = \sqrt
{\left( E_T-\frac{t_i}{2m_{0i}}\right) ^2-\left( 
p_T-\frac{E_T}{p_T}\frac{t_i}{2m_{0i}}-
\frac{t_i}{2p_T}\right) ^2-M_q^2} \simeq \sqrt{t_i} ,  
\end{eqnarray} 
the angle $\phi$ between the direction of vector ${\bf k_{ti}}$ and 
axis $z$ being distributed uniformly. The medium-induced radiative energy loss is 
calculated with Eq.~(\ref{rad_los}) without modification of the longitudinal rapidity. 

In our calculations, we use the Bjorken scaling 
solution~\cite{bjorken} for the 
space-time evolution of the energy density, temperature and 
density of the plasma:  
\begin{eqnarray} 
& & \varepsilon(\tau) \tau^{4/3} = \varepsilon_0 \tau_0^{4/3}, \\  
& & T(\tau) \tau^{1/3} = T_0 \tau_0^{1/3}, \\ 
& & \rho(\tau) \tau = \rho_0 \tau_0 .
\end{eqnarray}  
To be specific, we use the initial conditions for 
a gluon-dominated plasma expected for 
central Pb+Pb collisions at LHC~\cite{eskola94}: 
$\tau_0 \simeq 0.1$ fm$/c$, $T_0 
\simeq 1$ GeV, $N_f \approx 0$, $\rho_g \approx 1.95T^3$. 
It is interesting that the  
initial energy density, $\varepsilon_0$, in the dense zone 
depends on $b$ very slightly,
$\delta \varepsilon_0 / \varepsilon_0 \la 10 \%$, up to $b \sim R_A$ 
and decreases rapidly for  
$b \ga R_A$~\cite{lokhtin2}. On the other hand, 
the proper time of a jet to escape the dense zone 
averaged over all possible jet production vertices, $\left< \tau_L \right> $,
is found to decrease almost linearly with increasing impact parameter. This  
means that for impact parameters $b < R_A$, 
where $\approx 60 \%$ of the heavy quark 
pairs are produced~\cite{note00-060}, the difference in rescattering 
intensity and the corresponding energy loss is 
determined mainly by the different
path lengths rather than the initial energy density.  

The simulation of quark rescattering is halted if one of the following three
conditions is fulfilled: \\ 
1) A quark escapes from the dense zone, i.e. its path length becomes 
greater than the 
effective transverse spread of the matter from the production vertex 
to the escape 
point. The details of the geometrical calculations of these quantities 
at a given 
impact parameter can be found in Ref.~\cite{lokhtin2}. \\ 
2) The plasma cools down to $T_c=200$ MeV.  We thus neglect 
possible additional small contributions to the total energy loss due to 
re-interactions in the hadron gas. \\ 
3) A quark loses so much energy that its transverse momentum $p_T$ drops below
the average transverse momentum of the ``thermal" constituents of the medium.  
In this case, such a quark is considered to be ``thermalized" 
and its momentum in 
the rest frame of the fluid is generated from the random 
``thermal" distribution,  
$dN/d^3p \propto \exp{\left( -E/T\right) }$, boosted to the 
center-of-mass of the 
nucleus-nucleus collision~\cite{kamp98,lin98}. 

\section{High-mass dimuon production at LHC} 

Let us first consider dimuon production in the high invariant mass region, 
$20<M_{\mu^+\mu^-}<50$ GeV$/c^2$, where $$M_{\mu^+\mu^-} = 
\sqrt{(E_{\mu^+}+E_{\mu^-})^2 - ({\bf p}_{\mu^+}+{\bf p}_{\mu^-})^2}.$$ 
One of the main dimuon sources in this ``resonance-free" mass region is the 
semileptonic decays of open bottom and charm mesons. 
Heavy quark pairs are produced at 
the very beginning of the nuclear collisions by hard gluon-gluon scatterings 
and propagate through the dense medium. They finally form $B$ and $D$ mesons 
by ``capturing" $u$, $d$ or $s$ quarks during the 
hadronization stage. These mesons will decay with
the average meson lifetimes $c\tau_{B^{\pm}} = 496$ $\mu$m, 
$c\tau_{B^0} = 464$ $\mu$m, $c\tau_{D^{\pm}} = 315$ $\mu$m 
and $c\tau_{D^0} = 124$ $\mu$m.  We note that
$\approx 20 \%$ of $B$ mesons and $\approx 12 \%$ 
of $D$ mesons decay to muons. About 
half of the muons from $B$ decays are produced through an intermediate 
$D$~\cite{part_data} and contribute to the softer part of the $p_T$-spectrum. 
There is also dimuon production from single $B$ decays:  
$B\rightarrow \overline{D} \mu^+ X \rightarrow \mu^+\mu^-Y$. The 
branching ratio for this 
channel is comparable to the yield from $b \overline{b}$ pair decay. However 
the muon pairs from single $B$'s 
are concentrated in the low-mass region, $M_{\mu^+\mu^-} < M_B = 5.3$ 
GeV/$c^2$, below our interest here. 

Note that at LHC energies, there can be a significant contribution to heavy flavour 
production from gluon splittings, $g \rightarrow Q\overline{Q}$, in initial- or 
final-state shower evolution~\cite{norrbin}. Although the probability for a high-$p_T$ 
event to contain at least one $b \overline{b}$ or $c \overline{c}$ pair is fairly 
large, most of these quarks are carrying a small fraction of the total transverse 
momentum of the jet and dimuons produced in such a way are expected to be concentrated 
mainly in the low invariant mass region. Nevertheless the additional contribution of 
"showering" heavy quarks can be important even in high dimuon mass domain, and this 
phenomenon is studied in our next work~\cite{lok_new}. 

The main correlated background is Drell-Yan production, 
$q \overline{q} \rightarrow 
\mu^+\mu^-$. The uncorrelated part of the dimuon background, random decays of 
pions and kaons and muon pairs of mixed origin, is comparable 
with the signal from 
$b \overline{b}$-decays~\cite{note00-060} but these random decays also appear 
in the like-sign dimuon mass 
spectra. Thus such background can be estimated from the $\mu^+\mu^+$ 
and $\mu^-\mu^-$ 
event samples as  
\begin{equation}
\frac{dN^{\rm uncor}_{\mu^+\mu^-}}{dM} = 2 \sqrt{ \frac{dN_{\mu^+\mu^+}}{dM} 
\frac{dN_{\mu^-\mu^-}}{dM}} . 
\end{equation} 
and subtracted from the total $\mu^+\mu^-$ distribution. 
    
The cross sections, $\sigma _{NN}^{Q\overline{Q}}$, for 
heavy quark production in $NN$ 
collisions at $\sqrt{s}=5.5$ TeV, the initial $Q\overline{Q}$
momentum spectra, and the $B$ and $D$ meson 
fragmentation have all been obtained using PYTHIA$5.7$~\cite{pythia} with the  
default CTEQ2L parton distribution functions and including initial 
and final state 
radiation in vacuum which effectively simulates 
higher-order contributions
to heavy quark production. The corresponding Pb+Pb cross section is 
obtained by multiplying  $\sigma _{NN}^{Q\overline{Q}}$ by the 
number of binary nucleon-nucleon sub-collisions. The initial distribution of
$Q\overline{Q}$ pairs over impact parameter $b$ can be written 
as~\cite{note00-060,vogt99} 
\begin{equation} 
\label{jet_prob}
\frac{d^2 \sigma^0_{Q\overline{Q}}}{d^2b}(b,\sqrt{s})=T_{AA} (b)
\sigma _{NN}^{Q\overline{Q}} (\sqrt{s})
\frac{d^2 \sigma^{AA}_{\rm in}}{d^2b}(b,\sqrt{s})  
\end{equation} 
where the differential inelastic $AA$ cross section is   
\begin{equation} 
\frac {d^2 \sigma^{AA}_{\rm 
in}}{d^2b} (b, \sqrt{s}) = \left[ 1 - \left( 1- \frac{1}  
{A^2}T_{AA}(b) \sigma^{\rm in}_{NN} (\sqrt{s}) \right) ^{A^2} \right]   
\end{equation} 
and the total inelastic non-diffractive nucleon-nucleon cross section is 
$\sigma^{\rm in}_{NN} \simeq 60$ mb at $\sqrt{s} = 5.5$ TeV. The 
standard Wood-Saxon nuclear overlap function is 
$T_{AA}(b) = \int d^2s T_A (s) T_A (|{\bf b} - {\bf s}|)$ 
where $T_A(s) = A \int  dz \rho_A(s,z) $ is the 
nuclear thickness function with nucleon density distributions 
$\rho_A(s,z)$~\cite{ws}. 

We also take into account the modification of the nucleon structure functions due to 
the initial state nuclear interactions such as gluon depletion, nuclear shadowing, 
using the phenomenological parameterization of Eskola et al.~\cite{eks}. 
Figure 1 shows the initial $\mu ^+ \mu^-$ invariant mass spectra from 
correlated $b\overline{b}$, $c\overline{c}$ and Drell-Yan production 
in the CMS acceptance, $p_T^{\mu} > 5$ GeV/$c$ and $|\eta^{\mu}| < 2.4$. The upper and 
lower histograms for each
contribution show the results without and with nuclear shadowing respectively. 
In this kinematical region the influence of nuclear shadowing on heavy quarks
is relatively small, giving
an $\approx 15\%$ reduction in the open bottom and charm 
decays. Drell-Yan production is 
somewhat more affected, with an $\approx 25\%$ reduction, 
because quark shadowing is stronger 
than gluon shadowing at low $x$ in the EKS model~\cite{eks}. 
The total impact-parameter integrated 
rates are normalized to the expected number of Pb+Pb events 
during a two week LHC run, $R= 1.2 \times 10^6$ s, assuming luminosity 
$L = 10^{27}~$cm$^{-2}$s$^{-1}$~\cite{cms94} to that 
$$N(\mu^+\mu^-)= R \sigma^{\mu^+\mu^-}_{AA} L \, \, .$$ 

We see that the $b\overline{b}\rightarrow \mu^+\mu^-$ 
rates are greater than those
from the other sources by a factor of at least 5. 
Thus we only consider dimuons from $b\overline{b}$ 
decays in the remainder of the discussion. 
Moreover, the medium-induced charm quark energy loss 
can be significantly larger than the $b$-quark loss due to the mass 
difference, resulting in an additional suppression of 
the $c\overline{c}\rightarrow  \mu^+\mu^-$ yield. 

Drell-Yan pairs are unaffected by medium-induced final state interactions. 
These dimuons are directly from the primary nuclear interaction vertex 
while the dimuons from $B$ and $D$ meson decays appear at secondary vertices 
some distance from the primary vertex. The path length between the primary vertex 
and secondary vertices are determined by the lifetime and $\gamma$-factor of the 
mesons. This fact allows one cut to suppress the Drell-Yan rate by up to two orders 
of magnitude using the dimuon reconstruction algorithm based on the tracker information 
on secondary vertex position~\cite{note00-060}.

Figure 2 presents the  $\mu^+\mu^-$ invariant mass spectra from $b\overline{b}$ 
decays without and with the medium-induced $b$-quark energy loss
described in sections 2 and 3, including nuclear shadowing. The 
initial dimuon rate in the mass range $20<M_{\mu^+\mu^-}<50$ GeV/$c^2$ 
is $2.8 \times 10^{4}$ events per 
two week run without energy loss or shadowing. 
The rate can be reduced by factor of $1.6-4$ due to rescattering and 
$b$-quark energy loss in the QGP. Note that the rate integrated over 
all phase space is always conserved: the suppression in the rate 
appears only when kinematic cuts are specified. The dimuon suppression due to 
collisional loss is more pronounced at relatively lower invariant masses 
since the collisional energy loss is almost independent of the 
initial quark energy. The relative 
contribution of medium-induced gluon radiation to total energy loss grows with 
increasing $M_{\mu^+\mu^-}$ because the radiative energy loss grows 
with increasing initial energy, $\propto E$ in the 
incoherent limit used 
in our calculations, and thus contributes to the whole high-mass dimuon range, 
especially in the large invariant mass domain. The loss increases 
for high momentum quarks which 
contribute to the larger dimuon masses. 

Let us mark that the obtained value of dimuon rate suppression factor for the
scenario with collisional and radiative energy loss ($\sim 4$ at $M_{\mu^+\mu^-} \sim
20$ GeV) is close to the result of paper~\cite{lin98}, where a constant energy loss per 
unit length of $dE/dx = 1$ GeV$/$fm has been assumed. However, there exist the 
qualitative difference between invariant mass dependence of the effect in two models. 
In the case of constant (independent of energy) loss of heavy quark the suppression
factor in~\cite{lin98} decreases with increasing dimuon mass (this is similar to our 
scenario with collisional loss only), while the $M$-dependence in our model is 
specified by the dominant mechanism of energy loss. 

To conclude this section, we note that 
there are theoretical uncertainties in the 
bottom and charm production cross sections in nucleon-nucleon 
collisions at LHC energies which affect the quark production rates without
any loss effects. The absolute dimuon rates depend 
on the parton distribution functions, the heavy quark mass, 
the $B$-meson fragmentation scheme, next-to-leading 
order corrections, etc. Another possible contribution to 
the $b$ quark cross section could be light gluinos decaying into $b$ 
squarks~\cite{berger}, as proposed to explain 
the apparent deficit in the $B$ cross section 
at the Tevatron. It is therefore desirable 
that high mass dimuon measurements in $pp$ or $dd$ collisions are made
at the same or similar energy per nucleon as in the heavy ion runs.

\section{Dimuons from $B \rightarrow J/\psi$ at LHC}

We now consider another process which can also carry information about  
medium-induced rescattering and $b$-quark energy loss: secondary 
$J/\psi$ production. The branching ratio $B \rightarrow J/\psi X$ is $1.15\%$.
The $J/\psi$'s subsequently decay to dimuons with a $5.9\%$ branching 
ratio~\cite{part_data} so that e.g.  
$$gg \rightarrow b\overline{b}\rightarrow B\overline{B}~X\rightarrow 
J/\psi~Y\rightarrow\mu^+\mu^-Y .$$ 

Figure 3 shows the transverse momentum and pseudo-rapidity distributions of 
secondary $J/\psi$ decays for the same nuclear shadowing and energy 
loss scenarios described in the previous section. The total rates are 
again normalized to the expected 
number of events in a two week Pb+Pb run. We expect $1.3 \times 10^4$ dimuons
from secondary $J/\psi$'s at $M_{\mu^+\mu^-} = M_{J/\psi} = 3.1$ GeV/$c^2$. 
Including nuclear shadowing reduces this $J/\psi$ yield by $\sim 25\%$ while 
the final state rescattering 
and energy loss by $b$-quarks can further reduce the $J/\psi$ rates by a 
factor of $1.3-2.2$ in the CMS acceptance.  

We see that the influence of nuclear effects on secondary $J/\psi$ 
production and high-mass dimuon rates are quite different. The 25\% decrease 
of the secondary $J/\psi$ rate by nuclear 
shadowing is comparable to the $\sim 30-50\%$ effect of
medium-induced final state interactions.  On the other hand, the
high-mass dimuon rates are reduced by up to a factor of $4$ due to 
energy loss, much larger than 
the $15\%$ nuclear shadowing correction. The 
increased sensitivity to nuclear
shadowing is due to the different $x$ and $Q^2$ regions probed. 
The different influence of energy loss on secondary $J/\psi$ and 
high-mass dimuons is because secondary $J/\psi$'s come from the decay of a 
single $b$ quark instead of a $b \overline{b}$ pair and there is a
non-negligible probability that the energy lost by one quark is small. Thus a 
comparison between high-mass dimuon and secondary $J/\psi$ production could 
clarify the nature of energy loss. 

We note that about 5000 primary $J/\psi (\rightarrow \mu^+\mu^-)$ in a two week Pb+Pb 
run are expected to be initially produced by gluon-gluon fusion at CMS acceptance. The 
final primary $J/\psi$ rate is rather uncertain: on one hand, it should be suppressed 
due to colour screening~\cite{satz86} 
and/or dynamical dissociation in a QGP~\cite{khar94}; on the other hand, ``thermal" 
models predict some additional yield of $J/\psi$'s from a QGP, see, e.g. 
Refs.~\cite{vogt-psi} and references therein. Other models suggest
that$J/\psi$'s can be regenerated in the hadron phase by $D \overline D$
interactions \cite{redlich}.  Note that we call all these $J/\psi$'s 
``primary" in the sense that they are from the primary nuclear 
interaction vertex. Although they could be produced late in the 
collision, either after thermalization or in the hadronization stages, 
the time scale of their formation is much
less than the formation time of secondary $J/\psi$'s from $B$ meson decays.  
Thus it is necessary to distinguish secondary from primary 
$J/\psi$'s. As in the case of separating dimuons from Drell-Yan 
production and $b \overline{b}$ decays, primary $J/\psi$'s produced at the 
nuclear interaction point can be rejected 
using tracker information on the secondary vertex position. 
In addition, primary $J/\psi$'s have a softer dimuon $p_T$ 
spectrum than secondary $J/\psi$'s and the primary $J/\psi$ contribution disappears 
rapidly with increasing $p_T$. 

\section{Conclusions} 

To summarize, we have analyzed the sensitivity of spectra of high-mass 
$\mu^+\mu^-$ pairs from $b \overline{b}$ decays and dimuons from secondary 
$J/\psi$'s to medium-induced bottom quark energy loss in Pb+Pb collisions 
at LHC. For certainty the CMS experiment kinematical acceptance has been
considered. Since a complete description of coherent radiation
by massive quarks is still lacking, the dimuon spectra were 
calculated for two extreme cases: with collisional loss only, the ``minimum" effect, 
and with collisional and radiative loss estimated in the incoherent limit, 
the ``maximum" effect.  

We have found that medium-induced parton rescattering and energy loss can
reduce the dimuon rate in the invariant mass range 
$20<M_{\mu^+\mu^-}<50$ GeV/$c^2$ by a factor from $1.5$ to $4$ 
while nuclear shadowing is only of order of $15\%$ in 
our kinematical region. The relative contribution of radiative energy 
loss to the total dimuon suppression grows with increasing invariant 
mass and $p_T$ due to the 
stronger energy dependence of radiative loss relative to collisional loss. 

Since secondary $J/\psi$ production reflects the medium-induced energy loss of only 
one $b$-quark, the corresponding suppression by a factor of $1.3-2$ is 
less than for $b \overline{b}$ decays. On the other hand, 
the influence of nuclear shadowing on the relatively 
low invariant mass region, $M_{\mu^+\mu^-} \sim M_{J/\psi}$, seems to be 
non-negligible. We suggest that comparing 
high-mass dimuons with secondary $J/\psi$ 
production would help clarify the nature of this phenomenon. 

The experimental recognition of high-mass dimuons 
from $b \overline{b}$ decays relative to Drell-Yan 
pairs, as well as the recognition of secondary compared to primary 
$J/\psi$'s, could be performed using tracker information 
on the secondary vertex position. 

We conclude that the dimuon spectra will be sensitive to final state rescattering and 
energy loss of bottom quarks in dense matter. However, there are still 
theoretical uncertainties in the initial production of heavy flavours in 
nucleon-nucleon collisions at LHC energies. Thus measurements in $pp$ or 
$dd$ collisions at the same or similar energies per nucleon as in the heavy ion runs 
are required.\\[2ex]
 
{\it Acknowledgments}. 
It is a pleasure to thank D. Denegri for valuable suggestions and R. Vogt 
for important comments on this work. Discussions with M. Bedjidian, E. Boos, 
O.L. Kodolova, R. Kvatadze, L.I. Sarycheva, S. Shmatov and U. Wiedemann are gratefully 
acknowledged.

\begin{figure}
\centerline{\epsfig{figure=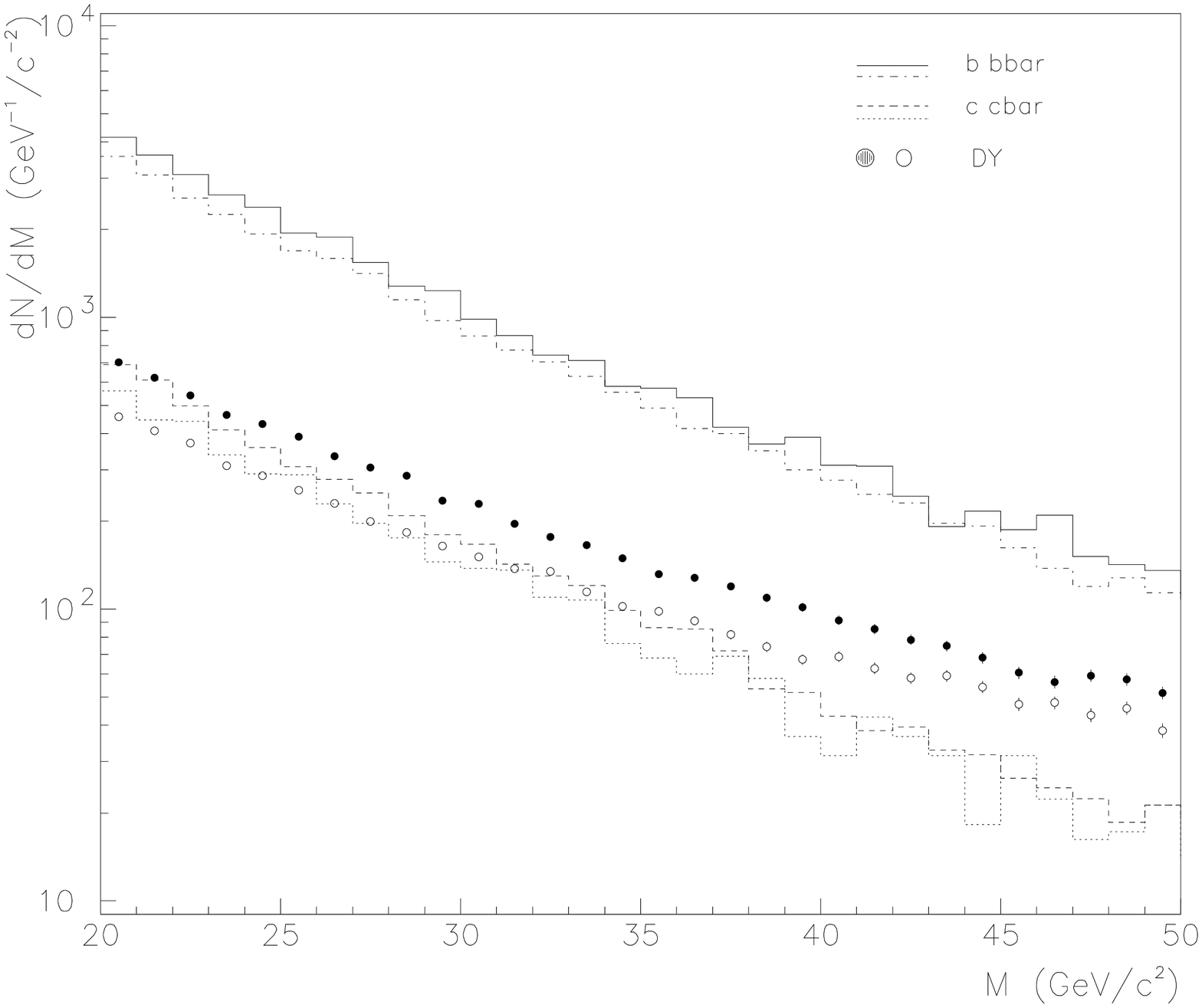,height=19cm}}
\caption{The initial invariant mass distribution of $\mu^+\mu^-$ pairs from 
correlated sources with $p_T^{\mu} > 5$ GeV/$c$ and $|\eta^{\mu}| < 2.4$. 
The results are for $b\overline{b}$ (solid and dot-dashed histograms -- without and 
with nuclear shadowing respectively), $c\overline{c}$ (dashed and dotted histograms), 
and Drell-Yan production (closed and open circles.)} 
\end{figure}

\begin{figure}
\centerline{\epsfig{figure=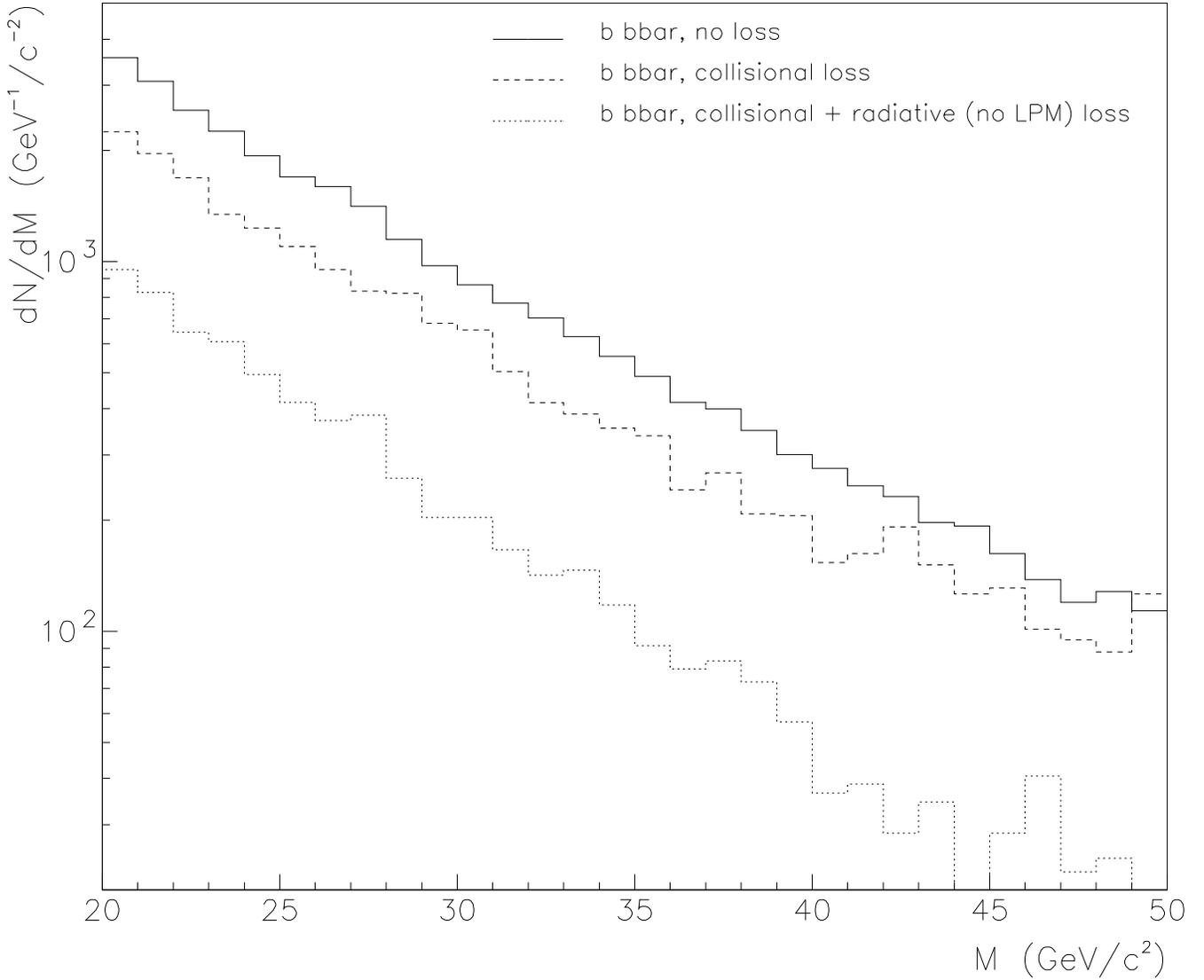,height=19cm}}
\caption{Invariant mass distribution of $\mu^+\mu^-$ pairs from 
$b \overline{b}$ decays with $p_T^{\mu} > 5$ GeV/c and $|\eta^{\mu}| < 2.4$ 
for various scenarios: without energy loss (solid histogram), with collisional loss 
only (dashed histogram), with collisional and radiative loss in incoherent limit 
(dotted histogram). Nuclear shadowing has been included.} 
\end{figure}

\begin{figure}
\centerline{\epsfig{figure=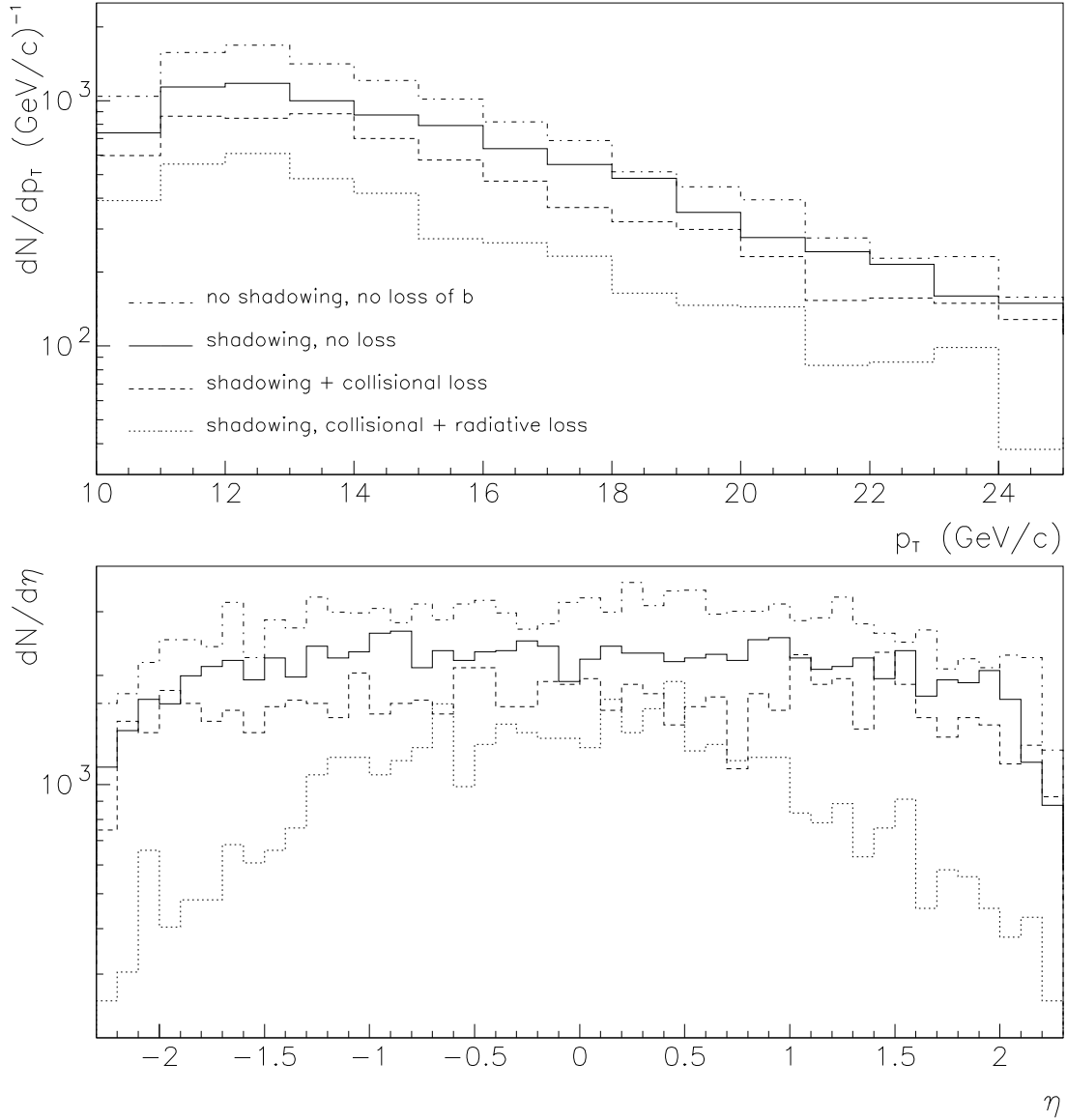,height=19cm}}
\caption{Transverse momentum and pseudo-rapidity distributions of secondary $J/\psi$ 
decays with $p_T^{\mu} > 5$ GeV$/c$ and $|\eta^{\mu}| < 2.4$ for various 
scenarios: without nuclear shadowing and energy loss (dash-dotted histograms), 
with shadowing and without loss (solid histogram), with shadowing and 
collisional loss (dashed histogram) and with shadowing, collisional and radiative loss 
in the incoherent limit (dotted histogram).} 
\end{figure}

\end{document}